\documentclass[onecolumn,amsmath,amssymb,prd,reprint,longbibliography,floatfix,8pt,a4paper,aps]{revtex4-1}

\usepackage{cancel}
\usepackage{enumitem}
\usepackage[utf8x]{inputenc}
\usepackage{graphicx}
\usepackage{dcolumn}
\usepackage{braket}
\usepackage{bm}
\usepackage{placeins}
\setcitestyle{super}
\usepackage[switch]{lineno}
\usepackage{float} 
\usepackage{subfigure}
\usepackage{tikz}
\usepackage{multirow}
\usetikzlibrary{arrows.meta}
\usetikzlibrary{arrows,decorations.pathmorphing,backgrounds,positioning,fit,petri,calc}
\allowdisplaybreaks

\begin{document}

\title{On the challenge of simulating dipolar contributions to spin relaxation with generalized cluster correlation expansion methods}

\author{Conor Ryan}
\author{Alessandro Lunghi}
\email{lunghia@tcd.ie}
\affiliation{School of Physics, AMBER and CRANN Institute, Trinity College, Dublin 2, Ireland}

\begin{abstract}
\noindent
The study of spin decoherence is often performed by assuming that spin-phonon interactions lead to relaxation at high temperatures, and spin-spin dipolar interactions instead contribute to pure dephasing at low temperatures. This has resulted in the neglect of spin relaxation due to spin-spin dipolar interactions and its influence on decoherence at low temperatures. For a complete understanding of low temperature spin dynamics, it is then imperative to focus also on the latter mechanism. One such method which has shown great promise in the efficient calculation of central spin dynamics due to spin-spin dipolar interactions with a surrounding spin bath is the Cluster-Correlation Expansion (CCE). An extension of this method through the explicit inclusion of the central spin degrees of freedom, known as the generalized Cluster-Correlation Expansion (gCCE) is capable of simulating the transfer of energy from the central spin into the bath, and thus could have the potential to investigate spin relaxation in this setting. In this work, we show that gCCE, in its standard form, is insufficient for providing even a qualitatively accurate description of spin-spin relaxation. A full mathematical deconstruction of the underlying theory of gCCE clearly points to the origin of such a breakdown and provides a starting point for its potential future resolution.
\end{abstract}

\maketitle

\section*{Introduction}
\noindent
The decoherence of central spin systems arising from interactions with an external spin bath has long been a topic of both theoretical and experimental interest.\cite{delgado2017spin, takahashi2011decoherence, wang2013spin, yao2006theory} This attention is driven not only by the desire to improve resolution in spin‑resonance spectroscopy,\cite{che1990electron, eaton2004electron} but also by the growing importance of spin systems as building blocks for emerging quantum technologies.\cite{coronado2020molecular, fursina2023toward, gaita2019molecular, kuppusamy2024spin, moreno2018molecular} A central quantity in this context is the coherence time $T_2$, which determines how long a spin can maintain quantum coherence, and therefore how long it can be resolved in spectroscopy, detect signals in quantum sensing, or undergo coherent operations in quantum computing.

In a dipolar spin bath, which constitutes a strongly non-Markovian environment,\cite{matern2019coherent} the central spin loses coherence through a combination of spin relaxation and spin dephasing processes. Unlike in a Markovian setting, these processes do not simply contribute independent additive rates,\cite{witzel2006quantum} as would be the case for a Markovian bath. Nevertheless, the overall coherence time remains bounded by the relaxation time $T_1$,
\begin{equation}
    T_2 \leq 2T_1.
\end{equation}
Spin relaxation processes are those in which the spin system of interest exchanges energy with an external bath, such as other spins or the vibrational modes of the crystal lattice. These processes in particular result in the finite lifetime of the excited spin state, or the excited component in a superposition of spin states. Dephasing processes, by contrast, involve no energy exchange between the spin and the bath and instead result in the loss of phase coherence among an identical ensemble of spins due to dynamical fluctuations in their environment. Static inhomogeneous contributions to $T_2$ are here neglected, as they can be easily removed with a simple Hahn-echo pulse sequence\cite{hahn1950spin} and do not pose particular challenges.

In the high temperature regime, typically already above 10-20 K, the dominant contributor to spin decoherence is spin relaxation due to spin-phonon interactions. This is the case independent of the spin system of interest, whether it be an NV center,\cite{bar2013solid, wood2022long} an impurity in P-doped Si,\cite{petersen2016nuclear, jeong2010spin} or a magnetic molecule.\cite{bader2014room, zadrozny2015millisecond, atzori2016room} At the microscopic level, this interaction arises from the modulation of the electronic structure defining the central spin by lattice vibrations and is made possible by relativistic interactions such as spin-spin and spin–orbit coupling. Having been the focus of intense research over the last few years, an in-depth understanding of how spin-phonon interactions, as a Markovian bath, contribute to both $T_1$\cite{lunghi2019phonons, garlatti2023critical, mondal2023spin} and a separate spin dephasing time $T_2^*$\cite{lunghi2023spin, lunghi2025fourth} has been recently developed.

In the low temperature regime, the temperature dependent phonon population rapidly decreases, and spin-phonon interactions become suppressed. In this temperature range $T_2$ is dominated by spin-spin dipolar interactions,\cite{witzel2012quantum} which arise due to magnetic dipolar interactions between the spin magnetic moment of the central spin and those of other spins present in the environment (the spin bath), be they identical spins to the central spin, or nuclear spins which are prominent in the lattice sites of the crystal, the ligands of magnetic molecules, and the solvent used to dilute the system.\cite{zadrozny2014multiple, ye2019spin}

For spin-spin dominated dynamics, the $T_1$ contributions to the decoherence of the central spin have only received limited attention for few select systems. Both the works of Mims et al.\cite{mims1960cross} and Yu et al.\cite{yu2020spin} have performed experimental investigations into these phenomena under the heading of ``cross relaxation''. The experiments of Mims et al. were performed not for magnetic molecules, but for synthetic ruby, and established a relationship between spin-spin dipolar relaxation and the concentration of the spin system. Those of Yu et al. then aimed to quantify the impact of spin-spin dipolar relaxation relative to phonon mediated processes in a bi-exponential decay, although lacked clarity that the fast process was indeed spin-spin dipolar relaxation rather than a faster phonon process. Ivady\cite{ivady2020longitudinal} and Cholsuk et al.\cite{cholsuk2025nuclear} have then approached the problem from a theoretical standpoint to calculate $T_1$ in NV centers and $V_B^-$ centers in hBN. While the predicted values of $T_1$ have been comparable to those in experiments, they have come at the cost of modeling the typically non-Markovian dipolar spin bath as Markovian, and opening the possibility of including empirical information in the calculations. Efforts must now be directed at establishing a fully quantitative theoretical understanding of spin-spin dipolar relaxation through an ab initio approach that reveals the underlying mechanisms of the process and allows the relaxation times to emerge from the microscopic physical principles. This theoretical understanding can then be used to directly inform the experimental conditions that allow for spin-dipolar mediated relaxation to truly be observed.

The performance of exact theoretical calculations in order to elicit any fundamental insight into spin-dipolar mediated relaxation would require simulation over an intractably large Hilbert space, due to its size scaling exponentially with the number of spins. To overcome the intractability of these calculations the Cluster-Correlation Expansion (CCE)\cite{yang2008quantum} method has been proposed. CCE has been successfully applied in the study of decoherence in a wide variety of spin systems, among which are solid-state defects such as NV centers and vacancies in hBN,\cite{nagura2025understanding, haykal2022decoherence} magnetic molecules,\cite{ryan2025spin, chen2020decoherence} and central spin systems consisting of entangled electron spins.\cite{chen2025simulating}

While the success of CCE in predicting $T_2$ has been well established across a host of spin platforms, particularly in the regime where relaxation is suppressed and decoherence is dominated by dephasing, there has been limited focus given to the explicit calculation of $T_1$ using this approach.\cite{yang2020longitudinal} Here we provide an explicit theoretical analysis on using the generalized Cluster-Correlation Expansion (gCCE), which explicitly includes the central spin degrees of freedom, to predict the relaxation times of spin systems as a result of dipolar interactions with their surrounding bath spins.

Our study demonstrates that the gCCE calculation of the central spin density matrix elements as a product of irreducible cluster contributions results in either unphysical dynamics, in which the population of spin state reaches values outside of [0,1], or overdamping of the relaxation. The consequence of the calculated dynamics in either case is the inaccurate calculation of the diagonal density matrix elements, and therefore an inability to provide precise quantitative predictions for $T_1$. The same mathematical framework as is used to highlight the issues of gCCE in the context of spin relaxation is then used to describe why gCCE is well-behaved for dephasing-dominated decoherence, thus completing a discussion on its validity as a numerical method.

\section*{Theoretical Background}
\noindent
In this section, the theory behind the spin-spin dipolar contributions to decoherence is explained, followed by a discussion on the general structure of the gCCE method as a way to tractably simulate the dynamics of spin open quantum systems.

\subsection*{Spin Relaxation and Dephasing}
\noindent
A general spin system containing a central spin $\vec{\textbf{S}}$, interacting with a spin bath $\{\vec{\textbf{I}}_i\}$ through spin-dipolar interactions is given by the following Hamiltonian,\cite{onizhuk2021pycce}
\begin{align}
    \hat{H} &= \vec{\textbf{S}}\cdot\textbf{D}\cdot\vec{\textbf{S}}+\vec{\textbf{B}}\cdot\gamma_S\cdot\vec{\textbf{S}}+\sum_i\vec{\textbf{S}}\cdot\textbf{A}_i\cdot\vec{\textbf{I}}_i+\sum_i\vec{\textbf{I}}_i\cdot\vec{\textbf{P}}_i\cdot\vec{\textbf{I}}_i\notag\\&+\vec{\textbf{B}}\cdot\gamma_i\cdot\vec{\textbf{I}}_i+\sum_{i<j}\vec{\textbf{I}}_i\cdot\textbf{J}_{ij}\cdot\vec{\textbf{I}}_j,
\end{align}
with $\textbf{D}(\textbf{P)}$ being the zero-field splitting (quadrupole) tensor of the central (bath) spin(s), $\textbf{A}$ the interaction tensor between the central and bath spins, $\textbf{J}$ the interaction tensor between bath spins, and $\gamma$ the interaction tensor between the spin and the magnetic field. Time evolution of an arbitrary spin system under this Hamiltonian will generally result in both relaxation and dephasing contributions to the decoherence of a particular spin within the system.

First characterizing the relaxation, $T_1$, contribution, in this work it refers broadly to processes where energy is transferred from the central spin into a surrounding spin bath, thus resulting in the decay of the central spin state to one of lower energy, as schematically depicted in Figure 1.
\begin{figure}[!h]
\centering
\begin{tikzpicture}[scale=0.5]
  \tikzset{
    level/.style={line width=1pt},
    solidarrow/.style={-{Latex[length=3mm]}, line width=1pt},
    dashedarrow/.style={-{Latex[length=3mm]}, dashed, line width=1pt},
    transition/.style={-{Latex[length=3mm]}, very thick},
    wavyarrow/.style={decorate, decoration={snake, amplitude=1.8pt, segment length=8pt}, -{Latex[length=3mm]}, line width=1pt}
  }

  \def\boxW{4.5}
  \def\boxH{4.5}
  \def\xL{-1.2}   
  \def\xR{1.2}
  \def\yTop{1.2}
  \def\yBot{-1.2}
  \def\arrowLen{0.9}
  \def\gap{2.2}   

  \draw[fill=green!20, draw=black, line width=1pt]
    (-\boxW/2, -\boxH/2) rectangle (\boxW/2, \boxH/2);
  \node[anchor=north west] at (-\boxW/2+0.2, \boxH/2-0.2) {$\mathbf{S}$};

  \draw[level] (\xL,\yTop) -- (\xR,\yTop);
  \draw[level] (\xL,\yBot) -- (\xR,\yBot);

  \draw[solidarrow] (0,\yTop-0.5) -- (0,\yTop+0.5);
  \draw[dashedarrow] (0,\yBot+0.5) -- (0,\yBot-0.5);

  \draw[transition] (0,\yTop) arc[start angle=90,end angle=270,radius=1.2];

  \draw[wavyarrow] (0,0) -- (\boxW/2+0.8,0)
    node[midway, above=6pt] {$\hbar\omega$};

  \begin{scope}[shift={( \boxW/2 + 0.8 + \gap , 0 )}]
    \draw[level] (\xL,\yTop) -- (\xR,\yTop);
    \draw[level] (\xL,\yBot) -- (\xR,\yBot);

    \draw[solidarrow] (0,\yBot+0.5) -- (0,\yBot-0.5);
    \draw[dashedarrow] (0,\yTop-0.5) -- (0,\yTop+0.5);

    \draw[transition] (0,\yBot) to[out=160,in=-160] (0,\yTop);
  \end{scope}

  \begin{scope}[shift={(-\boxW/2 - 1.5 , 1.8 )}]
    \draw[level] (\xL,\yTop) -- (\xR,\yTop);
    \draw[level] (\xL,\yBot) -- (\xR,\yBot);

    \draw[solidarrow] (0,\yTop-0.5) -- (0,\yTop+0.5);
  \end{scope}

  \begin{scope}[shift={(-\boxW/2 - 3.0 , -1.8 )}]
    \draw[level] (\xL,\yTop) -- (\xR,\yTop);
    \draw[level] (\xL,\yBot) -- (\xR,\yBot);

    \draw[solidarrow] (0,\yBot+0.5) -- (0,\yBot-0.5);
  \end{scope}

\end{tikzpicture}
\caption{\textbf{ Spin Relaxation.} A central spin undergoes relaxation with a spin bath by exchanging energy with resonant bath spins through spin flip-flops.}
\label{fig:spin_relaxation}
\end{figure}
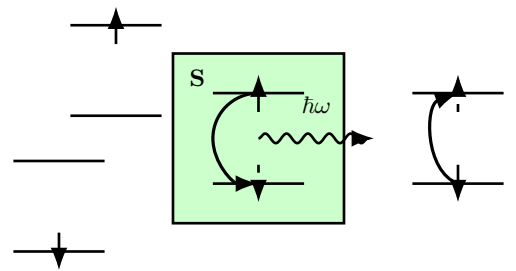
The explicit inclusion of the central spin operator, $\vec{\textbf{S}}$, in the Hamiltonian leads to the possibility of the $\bra{i}\hat{H}\ket{j}$ matrix element being non-zero, which in combination with the conservation of energy and selection rules being followed, will therefore enable transitions between the $\ket{i}$ and $\ket{j}$ states of the central spin. It is essential that the transfer of energy from the central spin into the spin bath conserve energy, hence bath spins in resonance with the central spin greatly enhance the process of energy transfer beyond the possibility of low probability spin flip-flops of out of resonance spins. For the central spin to relax it must therefore be immersed in a spin bath of an identical spin type, for example, an electron spin in a bath of electron spins, as they will all have similar Zeeman splittings and dipolar interactions with each other. The timescale of the transfer of energy from the central spin into the spin bath is characterized by the spin relaxation time, $T_1$.

For the pure dephasing contribution, there is no energy exchange between the central spin and the bath. The process is driven by energy-conserving spin flip-flops of bath spins which cause dynamical fluctuations of the magnetic field experienced by the central spin, and thus fluctuations in its Larmor frequency. These fluctuations then lead to a broadening of the central spin's spectral density, as illustrated in Figure 2.
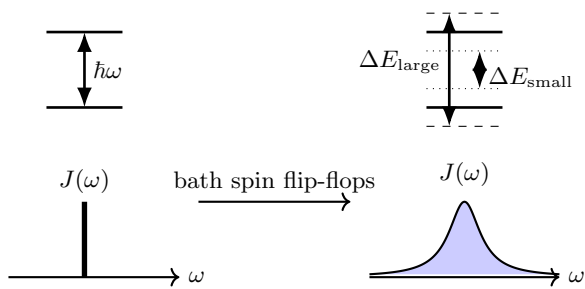
\begin{figure}[!h]
\centering
\begin{tikzpicture}[scale=0.5]
  \tikzset{
    axis/.style={->,thick},
    level/.style={line width=1pt},
    solidarrow/.style={Latex-Latex, line width=1pt},
    lorentz/.style={smooth, samples=100, domain=-2.5:2.5}
  }

  \begin{scope}[shift={(-5,-2)}]
    \draw[axis] (-2,0) -- (2.5,0) node[right] {$\omega$};
    \draw[line width=2pt] (0,0) -- (0,2.0);
    \node[above] at (0,2.0) {$J(\omega)$};
  \end{scope}

  \begin{scope}[shift={(5,-2)}]
    \draw[axis] (-2.5,0) -- (2.5,0) node[right] {$\omega$};
    \fill[blue!20, lorentz] plot (\x,{2/(1+(\x/0.5)^2)});
    \draw[thick, lorentz] plot (\x,{2/(1+(\x/0.5)^2)});
    \node[above] at (0,2.2) {$J(\omega)$};
  \end{scope}

  \begin{scope}[shift={(-5,3.5)}]
    \draw[level] (-1,1) -- (1,1);
    \draw[level] (-1,-1) -- (1,-1);
    \draw[solidarrow] (0,-1) -- (0,1);
    \node[right] at (0,0) {$\hbar\omega$};
  \end{scope}

  \begin{scope}[shift={(5,3.5)}]
    \draw[level] (-1,1) -- (1,1);
    \draw[level] (-1,-1) -- (1,-1);

    \draw[dashed] (-1,1.5) -- (1,1.5);
    \draw[dashed] (-1,-1.5) -- (1,-1.5);
    \draw[solidarrow] (-0.4,-1.5) -- (-0.4,1.5);
    \node[left] at (-0.4,0.2) {$\Delta E_{\text{large}}$};

    \draw[dotted] (-1,0.5) -- (1,0.5);
    \draw[dotted] (-1,-0.5) -- (1,-0.5);
    \draw[solidarrow] (0.4,-0.5) -- (0.4,0.5);
    \node[right] at (0.4,-0.2) {$\Delta E_{\text{small}}$};
  \end{scope}

  \draw[->,thick] (-2,0) -- (2,0) node[midway,above] {bath spin flip-flops};

\end{tikzpicture}
\caption{\textbf{ Spin Dephasing.} Spin flip-flops between the bath spins cause fluctuations in the magnetic field experienced by the central spin, which result in continuous shifts of its Zeeman levels. Its transition frequency then broadens from a well-defined value to a spectrum of possible values.}
\label{fig:spin_dephasing}
\end{figure}
It can be considered as the only contribution to $T_2$ in systems where the central and bath spins are far from resonance, such as an electron spin in a bath of nuclear spins, or its behavior isolated in systems that experience both relaxation and dephasing by projecting onto the central spin degrees of freedom. Performing this projection on the Hamiltonian in Eq (2) reduces it to
\begin{equation}
    \hat{H} = \sum_{i}|i\rangle\langle i|\otimes\hat{H}^{(i)},
\end{equation}
assuming $\bra{i}\hat{H}\ket{j}=0$, which is not true for any arbitrary spin system as relaxation could be possible, but is sufficient for isolating the dephasing contribution to decoherence. $\{\ket{i}\}$ are the eigenstates of the central spin, and the Hamiltonians $\{\hat{H}^{(i)}\}$ are conditioned on the state of the central spin and act only on the bath degrees of freedom.

The time evolution operator is then (assuming $\hbar=1$)
\begin{equation}
    e^{-i\hat{H}t} = \sum_{i}|i\rangle\langle i|\otimes e^{-i\hat{H}^{(i)}t}
\end{equation}
under this projection. For a central spin in an arbitrary superposition state $\ket{\psi}=\sum_ic_i\ket{i}$, and a spin bath in some pure state $\ket{\mathcal{J}}$, the combined system evolution is as follows,
\begin{equation}
    \sum_ic_i\ket{i}\otimes\ket{\mathcal{J}}\xrightarrow[]{e^{-i\hat{H}t}}\sum_ic_i\ket{i}\otimes\ket{\mathcal{J}_i(t)},
\end{equation}
where the bath state at any time will be conditional on the central spin state, $\ket{\mathcal{J}_i(t)}=e^{-i\hat{H}^{(i)}t}\ket{\mathcal{J}}$, meaning the overall state is an entangled state between the central spin and the spin bath.

As dephasing has no direct influence on the diagonal elements of the central spin density matrix, its effect can be quantified by analyzing the off-diagonal matrix elements, which can be calculated as the overlaps of the conditionally evolved bath states,
\begin{equation}
    \rho_{ij}(t) = \langle{\mathcal{J}_j(t)}|\mathcal{J}_i(t)\rangle = \bra{\mathcal{J}}e^{i\hat{H}^{(j)}t}e^{-i\hat{H}^{(i)}t}\ket{\mathcal{J}}.
\end{equation}
Information on the state of the central spin is then encoded in the evolution of the bath states, and as the evolution of these states is governed by different Hamiltonians, there will be a decay in their overlaps and, hence, the off-diagonal density matrix elements, known as the coherences.\cite{yang2008quantum}

\subsection*{Generalized Cluster Correlation Expansion (gCCE)}
\noindent
Exact time evolution of the closed central spin and spin bath system would require diagonalization of the full Hamiltonian given in Eq (2). A first approach that was developed to overcome this computational difficulty was to expand the propagators in the time evolution of the system,
\begin{equation}
    \rho(t) = e^{-i\hat{H}t}\rho(0)e^{i\hat{H}t},
\end{equation}
as a Dyson series.\cite{dyson1949s} From here, the arguments of perturbation theory can be applied and only the highest order terms in the expansion be retained in order to keep the terms which have the highest powers of the spin-bath coupling strength. In an analogous manner to the well-established methods in quantum field theory,\cite{PhysRev.76.769} each term in the perturbative expansion can be evaluated using a Feynman-diagram style approach known as the Linked-Cluster Expansion (LCE).\cite{saikin2007single} While LCE has been utilized to make accurate predictions for spin decoherence, the number of diagrams required for the calculations rapidly increases when considering higher orders in the spin-bath coupling strength and also for central spins with $S>1/2$. The LCE calculation process then quickly becomes tedious, and the need for a more efficient method evident.\cite{yang2008quantum}

Taking an arbitrary order in the perturbative expansion, such as order $k$, the corresponding term in the expansion contains\cite{saikin2007single}
\begin{equation}
    \int_0^tdt_1\dots\int_0^tdt_k\bra{n}\mathcal{T}\{\hat{H}_I(t_1)\dots\hat{H}_I(t_k)\}\ket{n},
\end{equation}
with $\{\ket{n}\}$ being pure states of the spin bath. The $k$ applications of the interaction picture Hamiltonian implies that this term contains interactions of clusters of bath spins up to size $k$. Taking this fact, combined with the physical intuition that on the timescale at which decoherence occurs it is possible for only relatively small sized clusters of bath spins to build up correlations through spin flip-flops, it is permissible to neglect dynamics in the spin bath which results in correlations between a relatively large number of bath spins. The perturbative expansion from LCE can therefore be rearranged so that the terms are ordered by the sizes of the interacting bath spin clusters instead of the order of the spin-bath coupling strength.\cite{yang2008quantum} The expansion of the central spin density matrix elements can then be written as a product of the irreducible correlations from all bath spin clusters due to spin flip-flops across all spins in the clusters,
\begin{equation}
    \rho_{ij}(t) = \prod_{|\mathcal{C}|}\tilde{\rho}_{ij,\mathcal{C}}(t).
\end{equation}
This expansion is known as the Cluster-Correlation Expansion (CCE).\cite{yang2008quantum} The irreducible cluster correlations are calculated by factoring out correlations of the subclusters from the dynamics of the selected cluster,
\begin{equation}
    \tilde{\rho}_{ij,\mathcal{C}}(t) = \frac{\rho_{ij,\mathcal{C}}(t)}{\prod_{\mathcal{C'}\subset\mathcal{C}}\tilde{\rho}_{ij,\mathcal{C'}}(t)},
\end{equation}
with the dynamics of the central spin and spin bath cluster being calculated unitarily as
\begin{align}
    \rho_{ij,\mathcal{C}}(t) &= \langle i|e^{-i\hat{H}_\mathcal{C}t}\rho_{S+\mathcal{C}}(0)e^{i\hat{H}_\mathcal{C}t}|j\rangle,
\end{align}
under the action of the cluster Hamiltonian\cite{onizhuk2021pycce}
\begin{align}
    \hat{H}_\mathcal{C} &= \vec{\textbf{S}}\cdot\textbf{D}\cdot\vec{\textbf{S}}+\vec{\textbf{B}}\cdot\gamma_S\cdot\vec{\textbf{S}}+\sum_{i\in\mathcal{C}}\vec{\textbf{I}}_i\cdot\textbf{P}_i\cdot\vec{\textbf{I}}_i+\vec{\textbf{B}}\cdot\gamma_i\cdot\vec{\textbf{I}}_i\notag\\&+\sum_{i<j\in\mathcal{C}}\vec{\textbf{I}}_i\cdot\textbf{J}_{ij}\cdot\vec{\textbf{I}}_j+\sum_{i\in\mathcal{C}}\vec{\textbf{S}}\cdot\textbf{A}_i\cdot\vec{\textbf{I}}_i+\sum_{a\notin\mathcal{C}}\vec{\textbf{S}}\cdot\textbf{A}_a\langle\vec{\textbf{I}}_a\rangle\notag\\&+\sum_{i\in\mathcal{C},a\notin\mathcal{C}}\vec{\textbf{I}}_i\cdot\textbf{J}_{ia}\langle\vec{\textbf{I}}_a\rangle.
\end{align}
This Hamiltonian consists of all terms for the central and cluster bath spins, with the effect of the non-cluster bath spins included in the last two terms from their mean-field average. These mean-field average terms then allow for the effect of the non-cluster bath spins to be included in a straightforward manner as a static field contribution to the dynamics. The explicit inclusion of the central spin degrees of freedom through the operator $\vec{\textbf{{S}}}$ in the Hamiltonian means that this approach in particular is known as the generalized Cluster-Correlation Expansion (gCCE), and in principle is capable of including the effect of both relaxation and dephasing processes in the dynamics of the central spin.

The physical picture of the CCE approach to make spin decoherence calculations tractable is to take a bath of $N$ spins and construct a set of overlapping clusters of the spins up to some set truncation size $M$, as visualized in Figure 3.
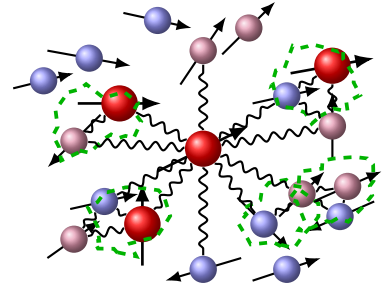
\begin{figure}[!h]
\centering
\begin{tikzpicture}[scale=0.5]

\tikzset{
  spinred/.style={ball color=red, circle, minimum size=0.48cm, inner sep=0pt},
  spinblue/.style={ball color=blue!40, circle, minimum size=0.36cm, inner sep=0pt},
  spinpurple/.style={ball color=purple!40, circle, minimum size=0.36cm, inner sep=0pt},
  interaction/.style={decorate, decoration={snake, amplitude=1.5pt, segment length=6pt}, thick},
  clusteroutline/.style={
    draw=green!70!black,
    line width=1.4pt,
    dashed,
    decorate,
    decoration={random steps, segment length=4pt, amplitude=3pt}
  }
}

\draw[-{Latex[length=2.5mm]}, line width=1pt] (-1.2,-0.6) -- (1.2,0.6);

\draw[-{Latex[length=2mm]}, line width=0.8pt] (1.3,1.2) -- (3.1,1.6);
\draw[-{Latex[length=2mm]}, line width=0.8pt] (3.4,-0.3) -- (3.4,1.5);
\draw[-{Latex[length=2.5mm]}, line width=1pt] (-3.3,1.2) -- (-1.1,1.2);
\draw[-{Latex[length=2mm]}, line width=0.8pt] (-2.7,0.9) -- (-4.1,-0.5);
\draw[-{Latex[length=2mm]}, line width=0.8pt] (0.9,-1.3) -- (2.3,-2.7);
\draw[-{Latex[length=2.5mm]}, line width=1pt] (-1.6,-3.1) -- (-1.6,-0.9);
\draw[-{Latex[length=2mm]}, line width=0.8pt] (1.0,-2.9) -- (-1.0,-3.5);
\draw[-{Latex[length=2mm]}, line width=0.8pt] (2.0,-1.8) -- (3.2,-0.6);
\draw[-{Latex[length=2mm]}, line width=0.8pt] (-3.5,-1.6) -- (-1.7,-1.2);
\draw[-{Latex[length=2mm]}, line width=0.8pt] (-0.6,1.7) -- (0.6,3.5);
\draw[-{Latex[length=2.5mm]}, line width=1pt] (2.3,2.0) -- (4.5,2.4);
\draw[-{Latex[length=2mm]}, line width=0.8pt] (-4.1,-2.9) -- (-2.7,-1.9);
\draw[-{Latex[length=2mm]}, line width=0.8pt] (1.3,-3.5) -- (3.1,-2.9);
\draw[-{Latex[length=2mm]}, line width=0.8pt] (-4.1,2.6) -- (-1.9,2.2);
\draw[-{Latex[length=2mm]}, line width=0.8pt] (0.5,2.5) -- (1.9,3.9);
\draw[-{Latex[length=2mm]}, line width=0.8pt] (-2.1,3.6) -- (-0.3,3.2);
\draw[-{Latex[length=2mm]}, line width=0.8pt] (2.9,-1.4) -- (4.7,-0.6);
\draw[-{Latex[length=2mm]}, line width=0.8pt] (4.5,-1.4) -- (2.7,-2.2);
\draw[-{Latex[length=2mm]}, line width=0.8pt] (-5.0,1.6) -- (-3.4,2.0);

\node[spinred] (central) at (0,0) {};

\node[spinblue]   (b1) at (2.2,1.4) {};
\node[spinpurple] (b2) at (3.4,0.6) {};
\node[spinred]    (b3) at (-2.2,1.2) {};
\node[spinpurple] (b4) at (-3.4,0.2) {};
\node[spinblue]   (b5) at (1.6,-2.0) {};
\node[spinred]    (b6) at (-1.6,-2.0) {};
\node[spinblue]   (b7) at (0,-3.2) {};
\node[spinpurple] (b8) at (2.6,-1.2) {};
\node[spinblue]   (b9) at (-2.6,-1.4) {};
\node[spinpurple] (b10) at (0,2.6) {};
\node[spinred]    (b11) at (3.4,2.2) {};
\node[spinpurple] (b12) at (-3.4,-2.4) {};
\node[spinblue]   (b13) at (2.2,-3.2) {};
\node[spinblue]   (b14) at (-3.0,2.4) {};
\node[spinpurple] (b15) at (1.2,3.2) {};
\node[spinblue]   (b16) at (-1.2,3.4) {};
\node[spinpurple] (b17) at (3.8,-1.0) {};
\node[spinblue]   (b18) at (3.6,-1.8) {};
\node[spinblue]   (b19) at (-4.2,1.8) {};


\draw[clusteroutline]
  plot [smooth cycle] coordinates {
    (2.05,1.15)
    (3.05,0.85)
    (3.75,1.25)
    (3.95,1.95)
    (3.55,2.55)
    (2.55,2.65)
    (1.95,2.05)
  };

\draw[clusteroutline]
  plot [smooth cycle] coordinates {
    (-3.75,0.05)
    (-3.25,-0.05)
    (-2.05,0.35)
    (-1.85,1.0)
    (-2.15,1.45)
    (-3.25,1.55)
    (-3.85,1.0)
  };

\draw[clusteroutline]
  plot [smooth cycle] coordinates {
    (-2.85,-2.45)
    (-2.25,-2.75)
    (-1.25,-2.35)
    (-0.95,-1.75)
    (-1.25,-1.15)
    (-2.05,-1.05)
    (-2.75,-1.45)
  };

\draw[clusteroutline]
  plot [smooth cycle] coordinates {
    (1.25,-2.35)
    (2.05,-2.55)
    (2.85,-2.0)
    (2.95,-1.45)
    (2.35,-1.0)
    (1.55,-1.15)
    (1.15,-1.75)
  };

\draw[clusteroutline]
  plot [smooth cycle] coordinates {
    (2.55,-2.15)
    (3.35,-1.95)
    (4.15,-1.55)
    (4.45,-0.85)
    (4.0,-0.25)
    (3.15,-0.35)
    (2.55,-1.05)
  };

\draw[interaction] (b1) -- (b2);
\draw[interaction] (b1) -- (b11);
\draw[interaction] (b2) -- (b11);

\draw[interaction] (b3) -- (b4);

\draw[interaction] (b6) -- (b9);
\draw[interaction] (b6) -- (b12);
\draw[interaction] (b9) -- (b12);

\draw[interaction] (b5) -- (b8);

\draw[interaction] (b8) -- (b17);
\draw[interaction] (b8) -- (b18);
\draw[interaction] (b17) -- (b18);

\foreach \s in {b1,b2,b3,b4,b5,b6,b7,b8,b9,b10}{
  \draw[interaction] (central) -- (\s);
}

\end{tikzpicture}
\caption{\textbf{ Generalized Cluster-Correlation Expansion (gCCE).} The central spin (center, red) interacts with a spin bath of different spin types (red, blue and purple) through spin-spin dipolar interactions (wavy lines). In the gCCE framework the bath spins are partitioned into clusters of different sizes, shown by the pairs and triplets of bath spins with the dashed outlines. Only the red bath spins will contribute to relaxation, as they are the same spin type as the central spin.}
\label{fig:gCCE}
\end{figure}
For CCE to be practically useful it is necessary that $M\ll N$, which is the case when CCE converges below maximum order. The approximate density matrix elements of the central spin calculated from CCE-$M$ are then
\begin{equation}
    \rho_{ij}^{(M)}(t) = \prod_{|\mathcal{C}|\leq M}\tilde{\rho}_{ij,\mathcal{C}}(t).
\end{equation}
By setting $M=N$, i.e. making the largest cluster be the entire bath, and expanding Eq (13) using Eq (10), the CCE calculation reduces to simulating the exact dynamics. This implies that CCE is guaranteed to converge by construction, although this limit will not be reached in systems of practical interest. Taking each bath spin to interact on average with $q$ other bath spins, with a typical pair flip-flop interaction strength of $\alpha_I$, a truncated CCE calculation will converge at times $T$ which satisfy $q\alpha_IT\ll1$.\cite{yang2008quantum} However, convergence at timescales beyond this range may still be possible.

In summary, CCE decomposes the dynamics of the central spin density matrix elements into a product of irreducible contributions from overlapping correlated bath spin clusters. On the timescale of interest, whether it be for the overall decoherence or specifically the relaxation component, the bath spin clusters are large enough to capture correlations from spin flip-flops involving all bath spins in the cluster, thus capturing the dynamically altering magnetic environment around the central spin, and also when using gCCE, the exchange of energy between the central spin and the bath spins. The problem is then reduced from simulating the dynamics of the very large Hilbert space of the full bath to one requiring the much more computationally tractable repeated simulation of reduced bath cluster Hilbert spaces.

\section*{Results}
\noindent
In this section, the fundamental theory and equations of gCCE, as given in the previous section, are discussed in more depth with a specific emphasis on their application in the problem of calculating the spin relaxation dynamics of a central spin.

\subsection*{Spin Relaxation}
\noindent
In the study of spin relaxation, the dynamics of interest is that of the diagonal density matrix elements of the central spin. Take the central spin to be initialized in a state $\rho_{++}(0)=1$. To introduce the specifics of the problem at hand, see the numerically simulated gCCE dynamics for a free electron central spin, i.e. not bound to any atomic nucleus, in a maximally mixed bath of 8 free electron spins presented in Figure 4. The specific quantity plotted in Figure 4 is $\rho_{++}^{(M)}(t)$ as given by Eq (13), for each CCE order $M$. The dynamics of the initial state density matrix element, $\rho_{++}$, is shown at all gCCE orders until it reaches the guaranteed exact dynamics at order 8.
\begin{figure}[!h]
    \centering
    \includegraphics[width=8cm]{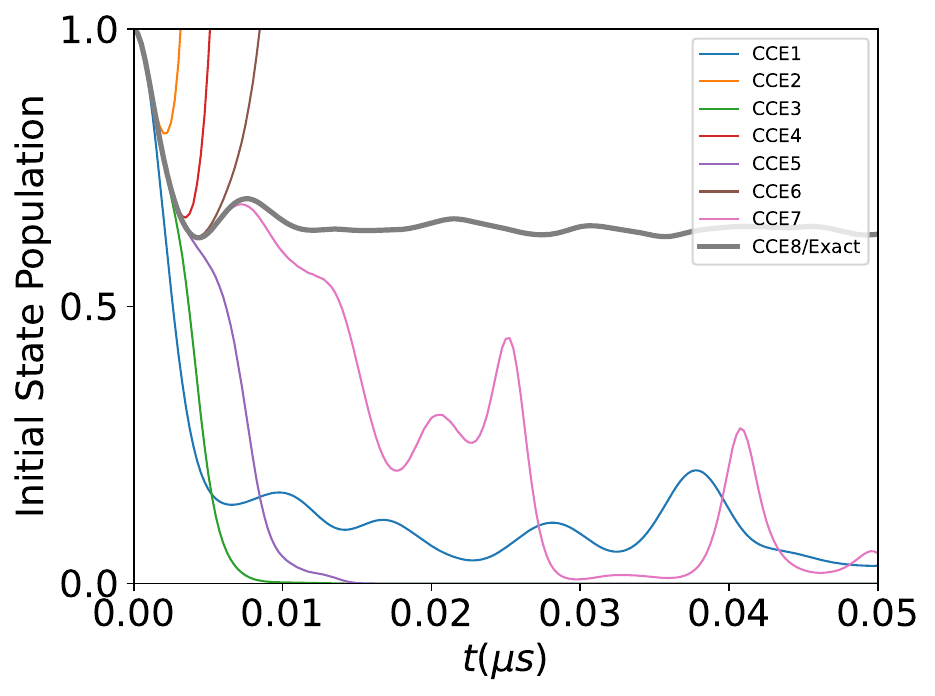}
    \label{fig:CCE_T1}
    \caption{\textbf{ CCE Relaxation Convergence.} Dynamical evolution of the spin-up state of a free electron interacting with a maximally mixed spin bath of 8 free electrons. The steady state of the dynamics either becomes unphysical or reaches an overdamped value where there is no longer any spin-up population.}
\end{figure}
Beginning with the exact result (gray line), it is seen that the central spin reaches an approximately thermal equilibrium state with the surrounding bath, retaining an initial state population of $\rho_{++}^{\text{eq}}\approx 0.63$. This is in direct line with expectations, as the excess initial population of $\rho_{++}(0)=1$ is distributed among the small number of bath spins in a relatively uniform manner. Figure 4 also shows that as the CCE order increases, which should lead the dynamics to convergence towards the exact result, there is no progression of the numerical results towards this point at any order below order 8. For orders below the exact result, as will always be the case in the simulation of any practical problem, the dynamics is either unphysical (CCE2,4,6), or tends towards a steady state where there is no longer any remaining population in the initial spin state (CCE1,3,5,7), which is not in any form of equilibrium with the surrounding spin bath. This lack of systematic numerical convergence provides our motivation for developing a quantitative analysis of the theoretical foundations of gCCE to assess its validity as a numerical method for simulating spin relaxation. To theoretically investigate the source of both the unphysical and overdamped dynamics, it is necessary to scrutinize the mathematical behavior demonstrated by the $\{\tilde{\rho}_{++,\mathcal{C}}(t)\}$ appearing in the product structure of CCE given by Eq (13), and thus their influence on the emergent value of $\rho_{++}^{(M)}(t)$.

Using the gCCE approach to simulate the relaxation of the central spin through spin-spin dipolar interactions, Eq (11) becomes
\begin{equation}
    \rho_{++,\mathcal{C}}(t) = \text{Tr}\left[\left(\rho_S(0)\otimes\rho_{\mathcal{C}}\right)\hat{U}^\dagger_\mathcal{C}(t)|+\rangle\langle+|\hat{U}_\mathcal{C}(t)\right].
\end{equation}
Eq (14) describes a density matrix element found from unitary evolution, meaning it will only ever predict physically allowed values, and in isolation, cannot be the source of the problematic dynamics. However, through Eq (10) it provides the first ingredient required to understand the behavior of the irreducible cluster contributions $\tilde{\rho}_{++,\mathcal{C}}(t)$. To begin developing this understanding, it is fruitful to take a deeper look into the dominant contributions to relaxation given by the time evolution in Eq (14). Expanding the propagators in Eq (14) in a Dyson series to 2nd order in the interaction picture Hamiltonian, $\hat{H}_\mathcal{C}(t)$, results in $\rho_{++,\mathcal{C}}(t)$ taking the following form at short times (a full derivation can be found in the Supplementary Material),
\begin{align}
    \rho_{++,\mathcal{C}}(t)&\approx1-\alpha_\mathcal{C}t^2+\mathcal{O}(t^3),\\
    \alpha_\mathcal{C}&=\text{Tr}\left[\rho\hat{H}_\mathcal{C}\hat{P}_\perp\hat{H}_\mathcal{C}\right],
\end{align}
where the short time approximation means that the time dependence of the interaction picture Hamiltonian can be dropped, and $\hat{P}_{\perp}$ is the projector onto the subspace of central spin states orthogonal to the initial state $\ket{+}$. For any state $\ket{\psi}$, $\bra{\psi}\hat{H}_\mathcal{C}\hat{P}_\perp\hat{H}_\mathcal{C}\ket{\psi}=\bra{\phi}\hat{P}_\perp\ket{\phi}\in[0,1]$ because $\hat{P}_\perp$ is a projector. Therefore $\alpha_\mathcal{C}\geq0$. The physical meaning of $\alpha_\mathcal{C}$ is that it quantifies the strength of the relaxation pathway provided by the cluster $\mathcal{C}$. A relaxation pathway can be taken as any collection of spin flip-flops between the central and bath spins resulting in the transfer of energy from the central spin into the spin bath. Eq (15) therefore shows that the leading order time evolution under the Hamiltonian in Eq (12) results in the dissipation of population from the initial $\rho_{++}(0)=1$ state due to the positivity of $\alpha_\mathcal{C}$. Thus, this expansion of $\rho_{++,\mathcal{C}}(t)$ is well-behaved as a diagonal density matrix element.

This result can now be utilized to write the irreducible cluster contributions, $\tilde{\rho}_{++,\mathcal{C}}(t)$, as
\begin{equation}
    \tilde{\rho}_{++,\mathcal{C}}(t) \approx 1-\alpha_\mathcal{C}^{irr}t^2 + \mathcal{O}(t^3).
\end{equation}
$\tilde{\rho}_{++,\mathcal{C}}(t)$ is not a density matrix element, and is therefore not obliged to obey the constraints given by the properties of density matrices, making its behavior more ambiguous. Insight into the leading order behavior of $\tilde{\rho}_{++,\mathcal{C}}(t)$ therefore requires an additional expression for $\alpha_\mathcal{C}^{irr}$. Inserting Eq (15) and Eq (17) into Eq (10) we obtain
\begin{equation}
    \tilde{\rho}_{++,\mathcal{C}}(t) \approx \frac{1-\alpha_\mathcal{C}t^2+\mathcal{O}(t^3)}{1-\sum_{\mathcal{C'\subset\mathcal{C}}}\alpha_\mathcal{C'}^{irr}t^2+\mathcal{O}(t^3)}.
\end{equation}
For sufficiently small $t$, a reciprocal may be expanded as follows using the geometric series identity,
\begin{equation}
    \frac{1}{1-At^2+\mathcal{O}(t^3)} = 1+At^2+\mathcal{O}(t^3),
\end{equation}
which simplifies Eq (18) as
\begin{align}
    \tilde{\rho}_{++,\mathcal{C}}(t) &\approx \left(1-\alpha_\mathcal{C}t^2\right)\left(1+\sum_{\mathcal{C'}\subset\mathcal{C}}\alpha_{\mathcal{C'}}^{irr}t^2\right)\notag\\
    &\approx 1-\alpha_\mathcal{C}t^2+\sum_{\mathcal{C'}\subset\mathcal{C}}\alpha_{\mathcal{C}}^{irr}t^2+\mathcal{O}(t^4)\notag\\
    &\approx 1 -\left(\alpha_\mathcal{C}-\sum_{\mathcal{C'}\subset\mathcal{C}}\alpha_{\mathcal{C'}}^{irr}\right)t^2+\mathcal{O}(t^4),
\end{align}
which when set equal to Eq (17) produces the following expression for $\alpha_\mathcal{C}^{irr}$,
\begin{align}
    \alpha_{\mathcal{C}}^{irr} &= \alpha_\mathcal{C}-\sum_{\mathcal{C'}\subset\mathcal{C}}\alpha_{\mathcal{C'}}^{irr}.
\end{align}
However, Eq (21) is recursive and still requires further work to define the $\{\alpha^{irr}_\mathcal{C}\}$ independently of each other. It is then useful to rearrange it to obtain
\begin{equation}
    \alpha_\mathcal{C} = \sum_{\mathcal{C'}\subseteq\mathcal{C}}\alpha^{irr}_\mathcal{C'}.
\end{equation}
Taking inspiration from combinatorics, this is exactly the standard M$\ddot{\text{o}}$bius–zeta relation on the poset of subsets ordered by inclusion.\cite{mobius1832besondere, rota1964foundations} We can then turn to the M$\ddot{\text{o}}$bius Inversion Theorem,\cite{mobius1832besondere, rota1964foundations} which states that for a function on a finite set of the form 
\begin{equation}
    f(\mathcal{C}) = \sum_{\mathcal{C'}\subseteq\mathcal{C}}g(\mathcal{C'}),
\end{equation}
it is also true that
\begin{equation}
    g(\mathcal{C}) = \sum_{\mathcal{C'}\subseteq\mathcal{C}}\mu(\mathcal{C'},\mathcal{C})f(\mathcal{C'}).
\end{equation}
On the poset of all subsets (subclusters) of some finite set (full cluster), ordered by inclusion, the M$\ddot{\text{o}}$bius function is\cite{mobius1832besondere, rota1964foundations}
\begin{equation}
    \mu(\mathcal{C'},\mathcal{C}) = \begin{cases}
    (-1)^{|\mathcal{C}|-|\mathcal{C'}|},\text{ if }\mathcal{C'}\subseteq\mathcal{C}\\
    0,\text{ otherwise}
    \end{cases}.
\end{equation}
The M$\ddot{\text{o}}$bius Inversion Theorem can therefore be applied directly to Eq (22) to obtain
\begin{equation}
    \alpha^{irr}_\mathcal{C} = \sum_{\mathcal{C'}\subseteq\mathcal{C}}(-1)^{|\mathcal{C}|-|\mathcal{C'}|}\alpha_\mathcal{C'},
\end{equation}
thus fully expressing $\alpha^{irr}_\mathcal{C}$ in terms of the $\{\alpha_\mathcal{C}\}$. The physical meaning of $\alpha_{\mathcal{C}}^{irr}$ is that it isolates the part of the relaxation pathway that requires all spins in the cluster $\mathcal{C}$. As Eq (26) makes clear, each subcluster $\mathcal{C'}\subseteq\mathcal{C}$ contributes a relaxation pathway $\alpha_\mathcal{C'}$, and the alternating signs subtract out every pathway that can already be generated by smaller subclusters. What remains is the genuinely new relaxation pathway that cannot occur unless all spins in $\mathcal{C}$ participate simultaneously.

To now investigate the numerical values that can be taken by $\tilde{\rho}_{++,\mathcal{C}}(t)$ and thus appear in the product in Eq (13), it is necessary to understand the mathematical behavior of $\alpha_\mathcal{C}^{irr}$. In general, $\alpha_\mathcal{C}^{irr}$ can be positive or negative because the sum in Eq (26) compares the relaxation pathway of the full cluster $\mathcal{C}$ against all relaxation pathways generated by its subclusters. If the collective relaxation pathway of $\mathcal{C}$ is more favorable than the alternating-sign combination of its subcluster pathways, the sum is positive; if the subclusters already provide more favorable relaxation routes, the alternating-sign subtraction outweighs $\alpha_\mathcal{C}$ and the irreducible term becomes negative.

This behavior can be intuitively understood by recognizing that clusters in the bath share spins, and therefore they also share relaxation pathways. Any two clusters that contain a common spin will include the same spin flip‑flop processes, leading to overlapping contributions to the central spin relaxation. If these overlaps were not removed, the same physical process would be counted multiple times. The alternating-sign structure of Eq (26) performs a systematic inclusion-exclusion cancellation of these shared pathways. Each subcluster $\mathcal{C'}\subset\mathcal{C}$ contributes the part of the dynamics that can already be explained by smaller groups of spins. By summing over all subclusters with alternating signs, the expansion removes every contribution that is reducible to lower-order processes. What remains is the irreducible contribution of the full cluster $\mathcal{C}$.

In the case where there is less overlap between the relaxation pathway of the full cluster $\mathcal{C}$ and those of its subclusters $\{\mathcal{C'}\}$, the relative magnitude of the pathway of the full cluster, $\alpha_\mathcal{C}$, will be larger than those of the $\{\alpha_{\mathcal{C'}}\}$. $\alpha_\mathcal{C}$ will then dominate the sum in Eq (26), resulting in $\alpha^{irr}_\mathcal{C}>0$ and the relaxation pathway of the full cluster $\mathcal{C}$ being more favorable than those of its subclusters. Alternatively, when the overlap between the relaxation pathway of the full cluster with the relaxation pathways of its subclusters is larger, it manifests quantitatively as larger numerical values of $\alpha_\mathcal{C'}$ because each subcluster contains many of the same spin‑flip processes that contribute to $\alpha_\mathcal{C}$, meaning the relative magnitude of $\alpha_\mathcal{C}$ in comparison to the $\{\alpha_\mathcal{C'}\}$ will be smaller. The sum in Eq (26) then produces $\alpha^{irr}_\mathcal{C}<0$ and the subclusters of $\mathcal{C}$ provide more favorable relaxation pathways than the one collectively involving all spins in $\mathcal{C}$.

First, taking the case in which there is strong overlap between the relaxation pathways, returning to Eq (17), $\alpha^{irr}_\mathcal{C}<0$ implies that $\tilde{\rho}_{++,\mathcal{C}}(t)>1$. For sufficiently many cluster overlaps, although without an exact quantitative idea of how many is sufficient, the product of many $\tilde{\rho}_{++,\mathcal{C}}(t)>1$ in Eq (13) leads to the prediction of unphysical values for $\rho_{++}^{(M)}(t)$. To demonstrate this effect, consider the same 9 electron system as used to produce Figure 4. The CCE2 result in Figure 4 is produced from Eq (13) by setting $M=2$ and considering the product over all possible bath spin clusters up to size 2 from the 8 bath spins. Instead, performing the product calculation in Eq (13) once again, but this time ignoring all order 1 clusters except for the clusters of the single spins \{1\},\{2\},\{3\}, and then only using order 2 clusters that do not include any of these spins, $\{4,5\},\{4,6\}\dots\{7,8\}$, the results become physical and convergent, as shown by the blue line in Figure 5.
\begin{figure}[!h]
    \centering
    \includegraphics[width=8cm]{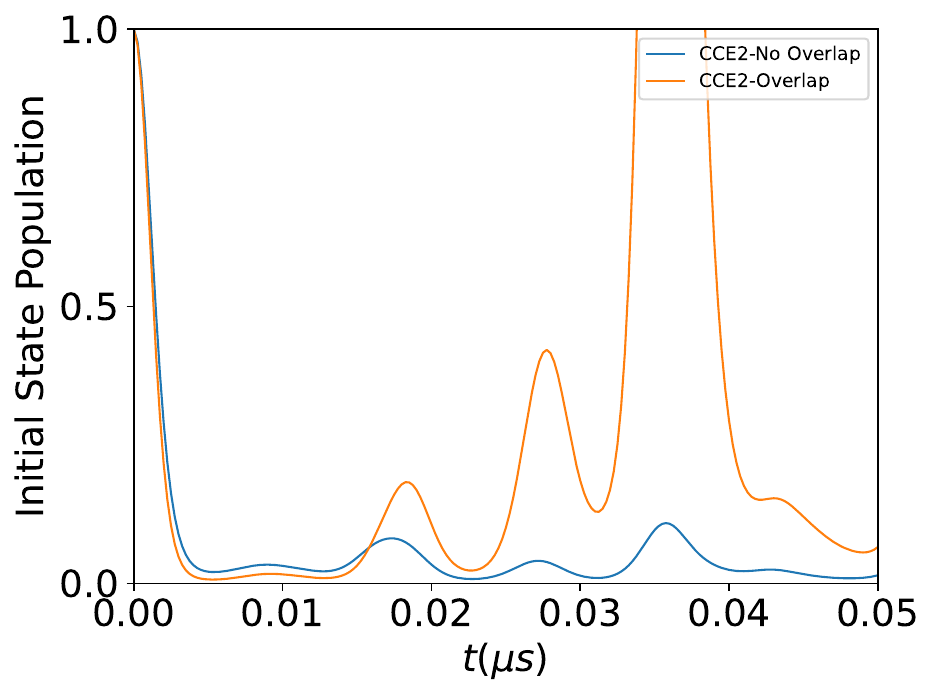}
    \label{fig:CCE_Overlap}
    \caption{\textbf{ CCE Convergence/Divergence.} Simulated dynamics for a 9 electron spin system at CCE order 2. Convergent behavior is calculated when there is no overlap between the order 2 clusters and the order 1 clusters (blue line), while the dynamics diverges once the order 2 clusters begin to overlap with the order 1 clusters (orange line).}
\end{figure}
Reintroducing all order 2 clusters containing spin 1, not even spins 2 or 3, then reintroduces overlaps between the order 1 cluster $\{1\}$ and the returning order 2 clusters $\{1,4\},\{1,5\},\{1,6\}\{1,7\}$ and $\{1,8\}$. The calculation from Eq (13) now returns to showing divergent behavior, as shown by the orange line in Figure 5. This result clearly points to the root cause of the instability of gCCE as a method being the strong overlap between the contributions of bath spin clusters.

With the theoretical cause of the unphysical dynamics now addressed, it remains to address the cause of the physical, but overdamped dynamics. Next, assume $\mathcal{C}$ contributes a more favorable relaxation pathway and therefore $\alpha^{irr}_\mathcal{C}>0$. This new relaxation pathway is represented by a dissipation of energy from the central spin involving collective transitions of all bath spins in the cluster. In this case, the irreducible cluster contributions $\tilde{\rho}_{++,\mathcal{C}}(t)$ are upper bounded by 1 for most of the bath spin clusters in the calculation. As a consequence of 1 being the upper bound of $\tilde{\rho}_{++,\mathcal{C}}(t)$ in the case of physical dynamics, the product in Eq (13) will be a product of a large number of values less than 1, resulting in the suppression of population recovery in the central spin and overdamping of the dynamics of $\rho_{++}^{(M)}(t)$ towards a steady state with no population in the initial spin state, exactly as seen in Figure 4. As time evolution for spin relaxation is described by quantum transition amplitudes, it follows that different clusters can generate the same final state of the central spin through their sharing of the same relaxation pathways. These contributions should then coherently interfere at the amplitude level to produce the resulting transition probability, much like in a superposition of states, the amplitudes of the states must be added before any probability can be calculated. This concept is for instance exemplified in the calculation of spin-phonon transition rates using the $T$-matrix approach\cite{lunghi2022toward}. In this formalism, the total rate is given by
\begin{equation}
    W_{i\rightarrow f} = \frac{2\pi}{\hbar}|\bra{f} \hat{T} \ket{i}|^2 \delta(E_f-E_i)\:,
\end{equation}
where the operator $\hat{T}=\sum_k \hat{T}^{(k)}$ runs over all possible $k$-phonon processes at the level of transition amplitudes prior to squaring to obtain the probability and finally taking a time-derivative to give the transition rate. This requirement is then in direct contradiction with the treatment of CCE where the probability of a spin state being occupied is calculated from a product of independent probabilities.

Having now analyzed Eq (26) for the meaningful cases when $\alpha^{irr}_\mathcal{C}>0$ or $\alpha^{irr}_\mathcal{C}<0$, as $\alpha^{irr}_\mathcal{C}=0$ only occurs when a cluster $\mathcal{C}$ does not contribute any relaxation pathway and is generally not found in practical situations, it is seen that the spin relaxation dynamics produced by gCCE can only be either unphysical or overdamped, neither of which provide any quantitative or even qualitative insight into the relevant phenomena.

\subsection*{Spin Dephasing}
\noindent
To create a complete picture of the validity of CCE as a numerical method through the mathematical framework employed in this work, a comparison must be made with its performance in the pure dephasing regime. In this case, Eq (13) will apply to a given off-diagonal density matrix element $\rho_{+-}(t)$. Once again taking the free electron system used in the previous examples, Figure 6 depicts how the calculation of $\rho_{+-}^{(M)}(t)$ for increasing $M$ does not represent the same numerical convergence issues as Figure 4 and converges towards the exact result. The dynamics shown is calculated from the average of 100 coherence profiles for the spin bath in a pure state selected through Monte Carlo sampling\cite{onizhuk2021pycce}. This successfully removes divergent behavior for coherence dynamics (see Discussion and Conclusions for more details), and allows for the clear depiction of the desired numerical convergence.
\begin{figure}[!h]
    \centering
    \includegraphics[width=8cm]{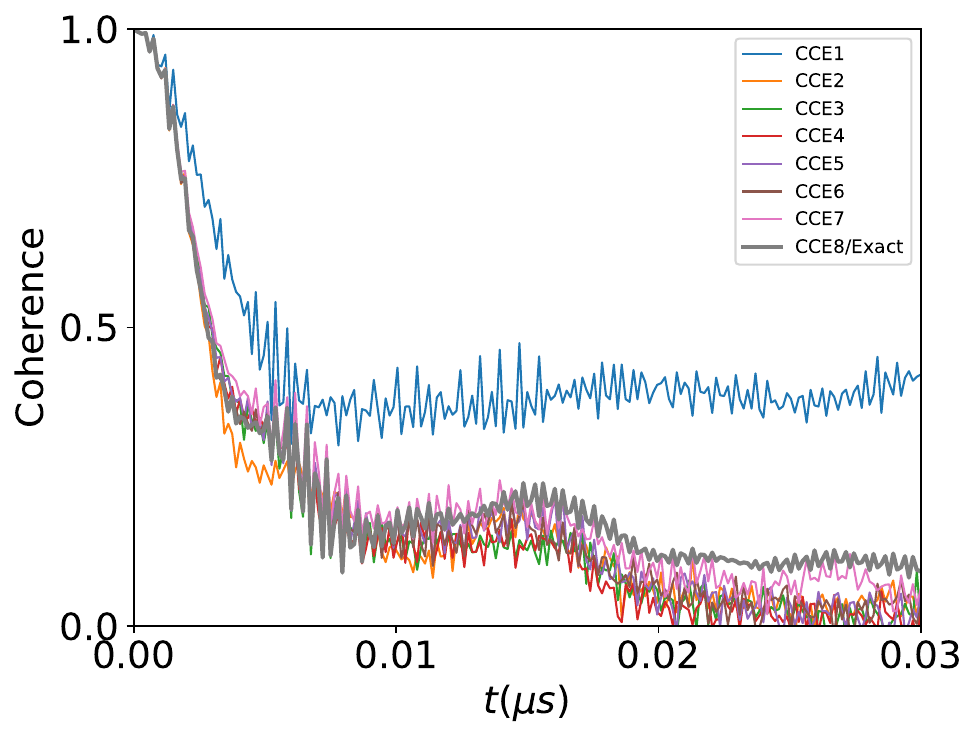}
    \label{fig:CCE_T2}
    \caption{\textbf{ CCE Dephasing Convergence.} Dynamical evolution of the coherence of a free electron interacting with a maximally mixed spin bath of 8 free electrons. The dynamics remains physical and systematically converges towards the exact result with increasing CCE order.}
\end{figure}

As the exact same equations of CCE apply as in the previous section, an identical time expansion can be performed for $\rho_{+-,\mathcal{C}}(t)$ through the same calculation process as in the derivation of Eq (15) and Eq (16), except replacing the projector $|+\rangle\langle+|$ with $|+\rangle\langle-|$ to obtain the desired density matrix element. The outcome of these calculations yields the following equations
\begin{align}
    \rho_{+-,\mathcal{C}}(t) &\approx1-\beta_\mathcal{C}t^2+\mathcal{O}(t^3),\\
    \beta_\mathcal{C}&=\frac{1}{2}\left(\langle\Delta\hat{H}_\mathcal{C}^2\rangle_{\rho_\mathcal{C}}-\langle\Delta\hat{H}_\mathcal{C}\rangle^2_{\rho_\mathcal{C}}\right),\\
    \Delta\hat{H}_\mathcal{C} &= \hat{H}_\mathcal{C}^{(+)}-\hat{H}_\mathcal{C}^{(-)}.
\end{align}
Where $\alpha_\mathcal{C}$ previously quantified the strength of the relaxation pathway contributed by the cluster $\mathcal{C}$, $\beta_\mathcal{C}$ now measures the strength of phase noise accumulated from the spin flip-flops of the cluster bath spins which results in the shifting of the central spin Zeeman levels. With analogously defined ingredients as for the spin relaxation case, the same arguments for analyzing the irreducible cluster contributions in Eq (13) apply and can be obtained through the same application of the M\"obius Inversion Theorem,
\begin{align}
    \tilde{\rho}_{+-,\mathcal{C}}(t) &\approx 1 - \beta_\mathcal{C}^{irr}t^2 + \mathcal{O}(t^3),\\
    \beta^{irr}_\mathcal{C} &= \sum_{\mathcal{C'}\subseteq\mathcal{C}}(-1)^{|\mathcal{C}|-|\mathcal{C'}|}\beta_\mathcal{C'}.
\end{align}
Having an identical set of equations with the exception of the definition of $\alpha_\mathcal{C}$ compared to $\beta_\mathcal{C}$, the difference in the relaxation and dephasing cases of gCCE now lies in how the physical meaning of the $\{\alpha^{irr}_\mathcal{C}\}$ makes spin relaxation incompatible with the multiplicative structure of CCE, whereas that of the $\{\beta^{irr}_\mathcal{C}\}$ does have this compatibility that enables accurate predictions of dephasing dynamics.

In an analogous manner to how $\alpha^{irr}_\mathcal{C}$ isolates the relaxation pathway arising from the collective spin flip-flops of all bath spins in the cluster $\mathcal{C}$, the identical structure of Eq (26) and Eq (32) ensures that $\beta_\mathcal{C}^{irr}$ isolates the irreducible phase fluctuations, i.e. fluctuations in the central spin Zeeman splitting generated by the simultaneous flip-flops of all bath spins in $\mathcal{C}$. The crucial distinction between the two cases emerges when these irreducible contributions are recombined through the product structure of Eq (13). For pure dephasing, Eq (32) shows that $\beta_\mathcal{C}^{irr}$ is built from phase-noise variances, the $\{\beta_\mathcal{C'}\}$ of Eq (29), which quantify the magnitude of the energy level fluctuations induced by each cluster. Taking the decay of the coherence as described by Eq (6), the time evolution of the bath state under different Hamiltonians results in the coherence acquiring a phase,
\begin{equation}
    \rho_{+-}(t)|_\mathcal{J} = e^{-i\phi_\mathcal{J}t}.
\end{equation}
A spin bath in a thermal state, rather than the pure state of Eq (6) then means the coherence can be written as\cite{redfield1957theory, witzel2006quantum}
\begin{align}
    \rho_{+-}(t) &= \sum_\mathcal{J}p_\mathcal{J}e^{-i\phi_\mathcal{J}t}\notag\\
    &= \langle e^{-i\phi_\mathcal{J}t}\rangle_\mathcal{J},
\end{align}
where, due to the independent effect of the clusters on the central spin, the phase fluctuations they induce add in the exponent and therefore combine multiplicatively at the level of the level of calculating $\rho_{+-}^{(M)}(t)$. Inserting Eq (31) into Eq (13) exactly reproduces this required multiplicative structure. In contrast, the relaxation contributions $\alpha^{irr}_\mathcal{C}$ are constructed from rate-like quantities, the $\{\alpha_\mathcal{C'}\}$ of Eq (16), which measure the strength of population transfer pathways between the central spin energy levels. Relaxation rates add rather than multiply, and thus inserting the irreducible relaxation terms of Eq (17) into the product of Eq (13) incorrectly treats distinct relaxation pathways as statistically independent multiplicative survival factors, leading to an overestimation of the total relaxation. Consequently, although the same mathematical inclusion-exclusion procedure isolates the correct irreducible physics in both cases, the product structure of the CCE is physically compatible only with dephasing, not with relaxation.

\section*{Discussion and Conclusions}
\noindent
The ability to make predictions on the behavior of complex systems is a requirement to further research to synthesize new materials and develop emerging quantum technologies. With a range of numerical methods being employed to carry out this task, it is imperative that their range of validity and limitations are well understood so that the predictions they provide can be trusted as reliable. Providing insight into the spin-spin dipolar interaction contribution to spin relaxation is one such problem that requires this predictive power.

In this work, we have taken the increasingly adopted CCE and gCCE methods and presented an analysis of their limitations and shortcomings in the context of being applied to the simulation of spin relaxation. We have contrasted these limitations with the widely demonstrated success that CCE has had when applied to decoherence problems dominated by dephasing. This has allowed us to highlight the precise theoretical details that mean gCCE, in its standard form, is unable to even qualitatively capture spin relaxation dynamics. It is our hope that this realization will stimulate research aimed at understanding if the computational benefits of CCE can be preserved while also overcoming these present limitations.

Our results clearly demonstrate that applying the CCE product structure to the population dynamics of the central spin produces one of overdamped relaxation, in which the initial-state population decays to zero rather than approaching a physically meaningful equilibrium with the spin bath, or unphysical dynamics, in which the population of the spin state attains values outside of [0,1]. This overdamped behavior arises because the CCE treats each bath cluster as providing an independent relaxation pathway that transfers population between the central-spin levels. The irreducible cluster contributions are then multiplied together, as in the standard CCE formulation, which implicitly assumes statistical independence of these pathways. However, relaxation pathways do not combine multiplicatively. In open quantum systems, population-transfer processes interfere and combine additively at the level of rates. The CCE product structure therefore overestimates the degree of relaxation of the central spin, leading to an exaggerated net relaxation rate and ultimately to overdamping. The inaccuracies caused by the independent treatment of the dynamics then lead to global quantities, such as energy, no longer being conserved, further underpinning the quantitative misrepresentation of the desired behavior.

A previous work by Yang et al.\cite{yang2020longitudinal} had made an attempt to directly apply gCCE to spin relaxation dynamics in NV centers. In their results they observed the problematic overdamped dynamics at the forefront of this work, although their attention was devoted towards discussing their prediction of spin relaxation on the qualitatively expected timescale. Our analysis now completes the explanation of their calculations.

The theoretical calculations of Ivady\cite{ivady2020longitudinal} and Cholsuk et al.\cite{cholsuk2025nuclear} were performed using an extended master equation approach in combination with CCE, evidently not encountering the problematic dynamics outlined in this work. However, where this approach gains in accuracy relative to the aforementioned shortcomings of standard gCCE, it makes a trade-off by invoking the Markovian approximation for the dipolar spin bath and opening the possibility of having to supply the calculations with empirical information about the noise rates of the spin bath. Thus, the promise of CCE as an ab initio approach to simulate spin dynamics, which also naturally captures the non-Markovian behavior of the spin bath, must be sacrificed in any case.

The comparison with the calculation of off-diagonal density matrix elements in the pure dephasing regime serves to validate the theoretical analysis used in the case of spin relaxation and reveals the underlying mathematical details that underpin CCE's success in this regime. Here, the product structure of CCE perfectly captures the required physical details of the dynamics. Spin flip-flops between bath spins each contribute independently to the phase fluctuations of the central spin, meaning the decomposition of the dynamics into independently contributing bath spin clusters as proposed by CCE exactly retains this feature and enables the product over all cluster contributions to build up the total accumulated phase fluctuations of the central spin and accurately predict its coherence dynamics.

For unphysical results, rather than systematically attempt to remove overlaps to reconverge the dynamics, which may not provide any guarantees on quantitative accuracy, divergent behavior has been suppressed in CCE through the Monte Carlo sampling of pure bath states.\cite{onizhuk2021pycce, ryan2025spin} In this approach, CCE is applied as normal, but time evolution is performed for $N$ sampled pure bath states from the probability distribution given by the actual thermal bath state, $\rho_B$, rather than directly using the thermal state itself. The dynamics given from the sampled states is then averaged out to accurately reproduce the desired bath statistics. For coherence dynamics, this approach is perfectly valid, as each bath state will produce dynamics with different phases that are static and additive, allowing them to faithfully provide accurate results after taking the average. However, for the spin relaxation dynamics at the focus of this work, any sampled pure bath state cannot capture the simultaneous possibility of a spin to relax or not relax that occurs with a thermal bath, thus resulting in the averaged dynamics failing to capture the behavior of the actual thermal spin bath. It is possible that the steady state of the averaged dynamics will be thermal, but this is more of a statistical artifact than any systematic correction of the calculations.

As the predictive power of CCE breaks down for population dynamics due to the product structure at the very heart of the method, one such way of bypassing these limitations would be to look elsewhere for methods which have quantitative accuracy and predictive power to provide insight into spin-dipolar mediated relaxation. Recently Li et al.\cite{li2025numerically} have applied tensor network methods to the dynamics of magnetic molecules. In this work, they have observed stable, non-diverging dynamics and demonstrate the potential for accurate steady states as well. Hierarchical Equations of Motion (HEOM)\cite{tanimura1989time, tanimura2020numerically} also offer a non-Markovian and non-perturbative approach that exactly treats the bath dynamics and could overcome the outlined issues discussed for CCE, although with a larger computational workload. There could also be promise in attempting to provide a modification of CCE that retains the computational efficiency and simplicity of its cluster form that has made it such a promising method, in which the incompatibility with the underlying physical structure of spin relaxation is overcome.

In conclusion, we have theoretically investigated the potential of simulating dipolar contributions to spin relaxation using the generalized Cluster-Correlation Expansion approach. Our results have demonstrated that, in its natural form, gCCE is unsuitable as a method to tackle this problem. The qualitative shortcomings of the gCCE predicted dynamics, namely unphysical dynamics or full population dissipation from the initial spin state and consequently a violation of energy conservation, prove to be critical issues in the calculation of $T_1$ as dictated by spin-spin dipolar interactions within this framework. The problem of understanding and making predictions on spin-dipolar mediated relaxation will thus require exploring alternative numerical recipes to open quantum systems dynamics or re-evaluate the core assumptions of CCE.

\section*{Supplementary Material}
\noindent
Detailed explanation of the derivation of Eq (15) and Eq (16) as presented in the manuscript.

\section*{Acknowledgments}
\noindent
A.L. acknowledges funding from the European Research Council (ERC) under the European Union’s Horizon 2020 research and innovation programme (grant agreement No. [948493]) and C.R. from Taighde Éireann- Research Ireland through the Government of Ireland Postgraduate Scholarship (project ID GOIPG/2024/4498). Computational resources were provided by Trinity College Research IT.

\section*{Author Contributions}
\noindent
A.L. conceived the project and supervised its execution. C.R. performed the simulations and theoretical analysis presented in this article. C.R. wrote the article with input from A.L.

\section*{Conflict of Interest}
\noindent
The authors declare that they have no competing interests.

\bibliographystyle{naturemag}
\bibliography{biblio}

\end{document}